\documentclass[final,onetabnum,onefignum,onethmnum]{siamltex}  
\usepackage{graphicx}

\usepackage[utf8]{inputenc}
\usepackage{color}
\usepackage{graphicx}
\usepackage[figure,noend,vlined,algo2e]{algorithm2e}
\usepackage{url}

\newcommand{\Minimum}{\mbox{$\mathit{minimum}$}}
\newcommand{\Maximum}{\mbox{$\mathit{maximum}$}}
\newcommand{\Insert}{\mbox{$\mathit{insert}$}}
\newcommand{\Extract}{\mbox{$\mathit{extract}$}}

\newcommand{\floors}[1]{\left\lfloor #1 \right\rfloor}
\newcommand{\ceils}[1]{\left\lceil #1 \right\rceil}
\newcommand{\set}[1]{\left\{#1\right\}}

\newcommand{\Access}{\mbox{$\mathit{access}$}}
\newcommand{\Rank}{\mbox{$\mathit{rank}$}}
\newcommand{\Select}{\mbox{$\mathit{select}$}}

\newcommand{\MinOne}{\mbox{$\mathit{min}\mbox{\rm -}\mathit{pile}_1$}}
\newcommand{\MinTwo}{\mbox{$\mathit{min}\mbox{\rm -}\mathit{pile}_2$}}
\newcommand{\MaxPile}{\mbox{$\mathit{max}\mbox{\rm -}\mathit{pile}$}}
\newcommand{\FirstCandidate}{\mbox{$\mathit{first}\mbox{\rm -}\mathit{candidate}$}}

\newcommand{\ylatyhjaa}{\mbox{${}^{\mbox{\rule{0cm}{1.25ex}}}$}}
\newcommand{\twodots}{\,.\,.\,}

\begin{document}

\title{Memory-Adjustable Navigation Piles\\{}with Applications to
  Sorting and Convex Hulls\thanks{Parts of this paper have appeared
  in preliminary form in \cite{AEK13} and \cite{DE14}}}

\author{%
Omar Darwish\thanks{Max-Planck Institute for Informatics, Saarbr\"{u}cken, Germany}
\and
Amr Elmasry\thanks{Department of Computer Engineering and Systems, Alexandria University, Egypt}
\and 
Jyrki Katajainen\thanks{Department of Computer Science, University of Copenhagen, Denmark}
}
 
\maketitle
\pagestyle{plain}
\pagenumbering{arabic}

\begin{abstract} 
We consider space-bounded computations on a random-access machine (RAM)
where the input is given on a read-only random-access medium, the
output is to be produced to a write-only sequential-access medium, and
the available workspace allows random reads and writes but is of
limited capacity.  The length of the input is $N$ elements, the length
of the output is limited by the computation, and the capacity of the
workspace is $O(S)$ bits for some predetermined parameter $S$.
We present a state-of-the-art priority queue---called an adjustable
navigation pile---for this restricted RAM model.  Under some
reasonable assumptions, our priority queue supports \Minimum{} and
\Insert{} in $O(1)$ worst-case time and \Extract{} in $O(N/S + \lg{}
S)$ worst-case time for any $S \geq \lg{} N$. (We use $\lg{} x$ as a
shorthand for $\log_2(\max\set{2, x})$.)
We show how to use this data structure to sort $N$ elements and to
compute the convex hull of $N$ points in the two-dimensional Euclidean
space in $O(N^2/S + N \lg{} S)$ worst-case time for any $S \geq \lg{} N$. 
Following a known lower bound for the space-time product of any branching program for finding unique elements, 
both our sorting and convex-hull algorithms are optimal.
The adjustable navigation pile has turned out to be useful when
designing other space-efficient algorithms, and we expect that it will find
its way to yet other applications.
\end{abstract}

\section{Introduction}
\label{sec:introduction}

\subsection{Problem Area}  Consider a sequential-access machine (Turing
machine) that has three tapes: input tape, output tape, and work
tape. In space-bounded computations, the input tape is read-only, the
output tape is write-only, and the aim is to limit the amount of space
used in the work tape. In this set-up, the theory of language
recognition and function computation requiring $O(\lg N)$ bits of
working space for an input of size $N$ is well established; people
talk about log-space programs \cite[Section 3.9.3]{Sav08} and classes
of problems that can be solved in log-space \cite[Section
  8.5.3]{Sav08}. Also, in this set-up, trade-offs between space and
time have been extensively studied \cite[Chapter 10]{Sav08}.  Although
one would seldom be forced to rely on a log-space program, it is still
theoretically interesting to know what can be accomplished when only a
logarithmic number of extra bits are available.

In this paper we reconsider the space-time trade-offs on a
random-access machine (RAM).  Analogous with the sequential-access
machine, we have separate storage media for read-only input, write-only
output, and read-write workspace which is of limited capacity.  Now,
however, both the input and workspace allow random access, but the
output is still to be produced sequentially.  Over the years, starting
by a seminal paper of Munro and Paterson \cite{MP80}---where a related
model was used, the space-time trade-offs in this \emph{restricted RAM
  model} have been studied for many problems including: sorting
\cite{Fre87,PR98}, selection \cite{EJKS14,Fre87}, and various
geometric problems \cite{ABBKMRS11,BKLSS15,DNR12,DNR15}.  The practical
motivation for some of the previous work has been the appearance of
special devices, where the size of working space is limited
(e.g.~mobile devices) and where writing is expensive (e.g.~flash
memories).

An algorithm or a data structure is said to be \emph{memory
  adjustable} if it uses $O(S)$ bits of working space for a given
parameter $S$.  Naturally, we expect to use at least a constant number of words, 
so $\Omega(w)$ is a lower bound for the space usage, $w$ being the size
of the machine word in bits. Sorting is one of the few problems for
which the optimal space-time product has been settled: Beame showed
\cite{Bea91} (see also \cite[Theorem 10.13.8]{Sav08}) that
$\Omega(N^2)$ is a lower bound, and Pagter and Rauhe showed
\cite{PR98} that an $O(N^2/S + N \lg S)$ worst-case running time is
achievable for any $S \geq \lg N$.
As for the convex-hull problem, Chan and Chen \cite{CC07} gave an
algorithm for computing the convex hull of a planar set of $N$ points,
stored in a read-only array, that runs in $O((N^2/S) \cdot \lg N + N
\lg S)$ worst-case time for any $S \geq \lg N$.

\subsection{Model of Computation} 
We assume that the elements being manipulated are on a read-only array, 
and use $N$ to denote the number of elements stored there.  
Observe that $N$ does \emph{not} need to be known beforehand.  
The output is sent to a separate write-only stream, 
where the output printed cannot be read or rewritten. 

In addition to the input and output media, a limited
random-access workspace is available.  The data on this workspace is
manipulated wordwise as on the word RAM \cite{Hag98}. We assume that
the word size $w$ is at least $\ceils{\lg N}$ bits and that the
processor is able to execute the same arithmetic, logical, and bitwise
operations as those supported by contemporary imperative programming
languages---like C \cite{KR88}.  It is a routine matter \cite[Section
  7.1.3]{Knu11} to store a bit vector of size $N$ such that it
occupies $\ceils{N/w}$ words and any string of at most $w$ bits can be
accessed in $O(1)$ worst-case time.  That is, the \emph{time
  complexity} of an algorithm is proportional to the number of
primitive operations plus the number of element accesses and element
comparisons performed.
We do \emph{not} assume the availability of any powerful memory-allocation
routines. The workspace is an infinite array (of words), 
and the space used by an algorithm is the prefix of this array. 
Even though this prefix can have some unused zones, the length of
the whole prefix specifies the \emph{space complexity} of the algorithm. 

In our setting, the elements lie in a read-only array and the data
structure only constitutes references to these elements. We assume
that each of the elements appears in the data structure at most once,
and it is the user's responsibility to make sure that this is the
case. Also, all operations are position-based; the position of an
element can be specified by its index.  Since the
positions can be used to distinguish the elements, we implicitly
assume that the elements are distinct.

\subsection{Our Results} 
Let $A$ be a read-only array and let $N$ denote its size.  Consider
a priority queue $Q$ storing a subset of the elements in $A$.  We use
$|Q|$ to denote the number of elements in $Q$. In the adjustable
set-up, a \emph{(min-)priority queue} is a data structure that
supports the following operations:
\begin{description}
\item[$Q.\Minimum{}()$:] Return the index of the minimum element
  in $Q$, as long as $|Q| > 0$.
\item[$Q.\Insert{}(i)$:] Insert  $A[i]$ into $Q$, for some $i
  \in\set{0,1,\ldots,N - 1}$.
\item[$Q.\Extract{}(i)$:] Extract $A[i]$ from $Q$, for some $i
  \in\set{0,1,\ldots,N - 1}$.
\end{description}
A \emph{max-priority queue}, which is defined to support the operation
$\Maximum{}$ instead of $\Minimum{}$, is obtained from a min-priority
queue by reversing the comparison function used in element
comparisons.
In the non-adjustable set-up, any priority queue---like a binary heap
\cite{Wil64} or a queue of pennants \cite{CMP88} (that both operate
in-place)---could be used to store positions of the elements instead
of the elements themselves.

In the first part of the paper, we improve and simplify the
memory-adjustable priority queue presented by Pagter and Rauhe
\cite{PR98} by introducing a kindred data structure that we call an
\emph{adjustable navigation pile}.  Compared to the navigation piles of
\cite{KV03}, that require $\Theta(N)$ bits, our adjustable variant can
achieve the same asymptotic run-time performance with only $\Theta(N/\lg N)$
bits. (Another priority queue that uses $\Theta(N)$ bits in addition
to the input was given in \cite{Elm03}.)  In Table
\ref{table:queues}, we compare the performance of the new data
structure to some of its competitors. Note that the stated bounds are
valid under some reasonable assumptions declared in Section \ref{sec:priority-queues}.

In the second part of the paper, we use the adjustable navigation pile
for sorting.  Our algorithm is simpler and more intuitive than that of
Pagter and Rauhe \cite{PR98}, and we also achieve the optimal
$O(N^2/S + N \lg{} S)$ running time for any $S \geq \lg{} N$.  
The algorithm is priority-queue sort like
heapsort \cite{Wil64}: Insert the $N$ elements one by one into a
priority queue and extract the minimum from that priority queue $N$
times.

In the third part of the paper, we improve the Chan-Chen bound for the
convex-hull problem by introducing an algorithm that runs in 
$O(N^2/S + N \lg S)$ time for any $S \geq \lg N$.  To prove this
result, we augment the adjustable navigation pile with extra
information while still using $O(S)$ bits of workspace.

\begin{table}[tb!]
\caption{The performance of adjustable navigation piles and their
  competitors in the restricted RAM model; $N$ is
  the size of the read-only input and $S$ is an asymptotic target for
  the size of workspace in bits where $S \geq \lg N$.}
\label{table:queues}
\begin{center}
\begin{small}
\tabcolsep0.75ex
\begin{tabular}{|c|c|c|c|c|}
\hline
\textbf{Reference}\ylatyhjaa & \textbf{Space\strut} & \Minimum{} &
\Insert{} & \Extract{}\\ 
\hline
\cite{CMP88}\ylatyhjaa{} & $\Theta(N \lg N)\strut$ & 
$O(1)$ & $O(1)$ & $O(\lg N)$\\
\cite{KV03} & $\Theta(N)$ & 
$O(1)$ & $O(\lg N)$ & $O(\lg N)$\\
\cite{Fre87} & $\Theta(S)$ & $O(1)$ & $O((N \lg N)/S +
\lg S)$ & $O((N\lg N)/S + \lg S)$\\
\cite{PR98} & $\Theta(S)$ &
$O(N/S^2 + \lg S)$ & $O(N/S + \lg S)$ amortized & $O(N/S + \lg^2 S)$\\
{}[this paper] & $\Theta(S)$ &
$O(1)$ & $O(1)$ & $O(N/S + \lg S)$\\
\hline
\end{tabular}
\end{small}
\end{center}
\end{table}

\subsection{Related Models} The basic feature that distinguishes the
model we use from other related models is the capability of having
random access to the input data. In the context of sequential-access
machines, the input is on a tape that only allows single-pass
algorithms. The so-called \emph{streaming model} still enforces
sequential access, but allows multi-pass algorithms, and the goal is
to minimize the number of passes over the input when the size of the
random-access workspace is limited. Munro and Paterson \cite{MP80}
considered this model. For some problems, the restricted RAM model is
more powerful than the multi-pass streaming model. For example, for
the selection problem the lower bound known for the multi-pass
streaming model \cite{Cha10} can be bypassed in the restricted RAM
model \cite{EJKS14}.

In the \emph{in-place model}, the elements are to be stored at the
beginning of an infinite array, a constant number of additional
variables are allowed, and the elements may be swapped and
overwritten, but not modified, still keeping this compact
representation.  All the problems considered in this
paper---maintaining priority queues (see,
e.g.~\cite{CMP88,EEK15,Wil64}), sorting (see,
e.g.~\cite{EEK15,KPT96,Wil64}), and computing convex hulls
\cite{BIKMMT04}---have been studied in this classical setting.  In the
\emph{restore model}~\cite{CMR14}, the input elements may be
temporarily rearranged, and even modified, during a computation, but
at the end the input must be restored to its original state. Again,
sorting problems have been central in the exploration of the power of
this model (see \cite{CMR14,KP94}). In general, any reversible
algorithm would work well in the restore model, so reversible
computing is distantly related to this study.

\subsection{Bit Vectors with \Rank{} and \Select{} Support}
Given a bit vector $B$ of $N$ bits, of which $n$ are $1$ bits, consider the
following operations:

\begin{description}
\item[$B.\Access(i)$:] Return $B[i]$, i.e.~the bit at index $i$ for
  some $i \in\set{0,1,\ldots,N-1}$.
\item[$B.\Rank(i)$:] Return the number of $1$ bits among the bits
  $B[0], B[1],\ldots, B[i]$, for some $i \in\set{0,1,\ldots,N-1}$.
\item[$B.\Select(j)$:] Return the index of the $j$th 1 bit, for some $j \in\set{1,2,\ldots,n}$, i.e.~if
  $B.\Select(j) = i$, this means that $B[i]=1$ and $B.\Rank(i) =
  j$. 
\end{description}
The operations $B.\Rank\mbox{0}(i)$ and $B.\Select\mbox{0}(j)$ are
similarly defined considering the $0$ bits instead of the $1$ bits.

For a bit vector of size $N$, our requirements for an acceptable
solution are that all the operations should run in
$O(1)$ worst-case time, the space used should be $O(N)$ bits,
and the construction of the data structure should take $O(N)$
worst-case time.  The problem of extending a bit vector with
\Rank{}-\Select{} operations has been addressed in several papers
(see, for example, \cite{Jac89,Mun96,RRR07}).  Most of the known
solutions rely on the idea of dividing the bit vector into blocks,
precomputing \Rank{} and \Select{} values for some specific
positions, and calculating the other values on the fly using the
stored values, some precomputed tables, and bits in the bit vector
itself.  For example, the solution presented in \cite{RRR07} would be
suitable for our purposes; it requires $N + O(N\lg\lg N/\lg N)$
bits, and $O(1)$ worst-case time per operation.

\section{Memory-Adjustable Priority Queues}
\label{sec:priority-queues}

\subsection{Assumptions}
In this section two memory-adjustable priority queues are described.
The first structure is a straightforward adaptation of a tournament tree
(also called a selection tree \cite[Section 5.4.1]{Knu73}) for
read-only data. For a parameter $S$, it uses $O(S)$ words of workspace. 
The second structure is an improvement of a navigation pile \cite{KV03} for which 
the workspace is $O(S)$ bits, for $S \geq \lg N$, where $N$ is the size of the 
read-only input. Both data structures can perform \Minimum{} and \Insert{} 
in $O(1)$ worst-case time and \Extract{} in $O(N/S + \lg S)$ worst-case time.

When describing the data structures, we
make the following assumptions:
\begin{enumerate}
\item $N$ is known beforehand.
\item Insertions and extractions are \emph{monotonic} such that,
  at any given point of time, there is a single element (if any)
  indicating that the elements smaller than or equal to it are outside
  the data structure.  We call such an element the \emph{latest
    output}, and say that an element is \emph{alive} if it is larger
  than the latest output. In particular, an extraction must remove the
  smallest element and an insertion must add an element that is larger
  than the latest output.
\item Insertions are \emph{sequential}---but insertions and
  extractions can be intermixed---such that the elements are inserted
  one by one from consecutive input entries starting from the first
  element stored in the read-only input.
\end{enumerate}
These assumptions are valid when a priority queue is used
for sorting. Actually, in sorting all insertions are executed before
extractions; a restriction that is not mandated by the data structure.
At the end of this section, we show how to get rid of these
assumptions. The first assumption is not critical. But, 
when relaxing the second assumption, the required size of workspace has to increase by $N$ bits.
When relaxing the third assumption, the asymptotic worst-case running time of \Insert{}
will become the same as that required by \Extract{}.

\subsection{Adjustable Tournament Trees} 
For an integer $S$, we use $\bar{S}$ as a
shorthand for $2^{\ceils{\lg S}}$. It suffices that $S \leq N/\lg N$; even if $S$ was larger, the
operations would not be asymptotically faster. The input array is
divided into $\bar{S}$ \emph{buckets} 
each containing $\lceil N/\bar{S} \rceil$ elements, except for the last bucket that may contain less.
A complete binary tree is built above these buckets. Each leaf of
this tree
\emph{covers} a single bucket and each branch covers the
buckets of the leaves in the subtree rooted at that branch. We
call the elements within the buckets covered by a node the
\emph{covered range} of this node. Note that the covered range of a
node is a sequence of elements stored in consecutive locations of the
input array. The data stored at each node is an
index specifying the position of the smallest alive element in the
covered range of that node.
 
\begin{figure}[tb!]
\begin{center}
\includegraphics[scale=0.41]{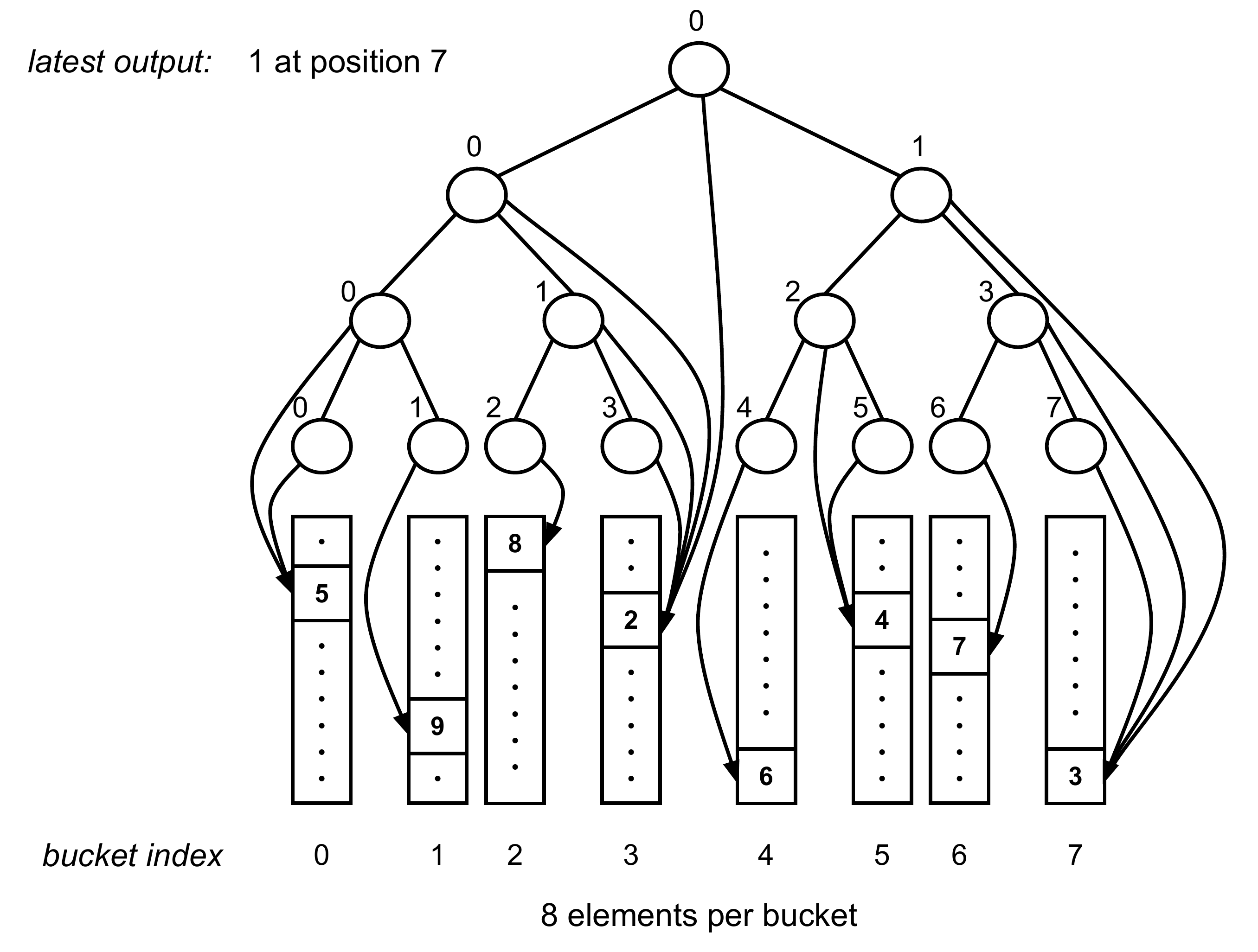}
\end{center}

\caption{An adjustable tournament tree for a set of $N= 64$ elements
  when $\bar{S} = 8$. Indices stored at the nodes are visualized as
  pointers. Only the smallest alive element in each bucket is shown.}
\label{fig:tournament}
\end{figure}

An \emph{adjustable tournament tree} is an array of $2 \bar{S} - 1$ positions (indices). 
To make the connection to our adjustable navigation piles clear, 
we store the indices in this array in breadth-first order as in a binary heap \cite{Wil64}. 
At each level, we start the indexing of the nodes  from
0. For the sake of simplicity, we maintain a \emph{header} that stores
the offsets to the beginning of each level (even though this
information is easy to calculate). For a node number $i$, its
left child has number $2i$ at the level below, the right child has number $2i +1$ at the level below, 
and the parent has number $\floors{i/2}$ at the level above. When we know the current level and the index
of a node at that level, the information available at the header and the above formulas
are enough to get to a neighbouring node in constant time. In
Figure~\ref{fig:tournament}, we give an illustration of an adjustable tournament tree
for a set of $N=64$ elements when $\bar{S} = 8$.

To support \Insert{} efficiently, we partition the data structure into
three components: tournament tree, submersion buffer, and
insertion buffer.  The \emph{submersion buffer} is the last full
bucket that is being submerged into the tournament tree. The
\emph{insertion buffer} is the bucket that embraces the new
elements. One or both of these buffers can be empty. The
idea is to insert the elements into a buffer and, when the buffer gets
full, submerge it into the tournament tree.  To make \Minimum{} 
straightforward, we recall the position of the overall
minimum of these three components.

In Figure~\ref{fig:submersion}, we describe in pseudo-code how a
bucket is submerged into the tournament tree at one go. In \Insert{},
the submersion is done incrementally. The process starts from a leaf
corresponding to the given bucket and proceeds in a bottom-up manner
to the root by following the implicit parent pointers.  We call the
path of nodes visited the \emph{updating path}. First, the index at
the leaf is set to point to the minimum of the bucket. Then, every
branch node on the updating path inherits the position of the smaller
of the two elements pointed to via its two children.  At the end the
root stores the index of the overall minimum among all alive
elements. Since, at some point, some buckets may have no alive elements, 
we use an unspecified constant $\mathbf{none}$ to indicate that the  
covered range of the node in question has no alive elements.

\begin{algorithm2e}[tb!]
\begin{minipage}{1.1\textwidth}
\SetStartEndCondition{ }{}{}%
\SetKwInput{Proc}{procedure}
\SetKwInput{Input}{input}
\SetKwInput{Data}{data}  
\SetKwFor{For}{for}{{\rm :}}{}
\SetKwIF{If}{ElseIf}{Else}{if}{{\rm :}}{else if}{else{\rm :}}{}%
\DontPrintSemicolon
\Proc{$\mathit{submersion}$}
\Input{$\mathit{bucket}${\rm -}$\mathit{start}$: index of the
  beginning of the bucket to be submerged\newline
$\mathit{bucket}${\rm -}$\mathit{min}$: index of the minimum alive element
  within this bucket} 
\Data{$N$: number of elements; $\bar{S}$:
  workspace target rounded to a power of 2\newline
$A[0\twodots N-1]$: read-only array of elements\newline $T[0\twodots 2\bar{S}
    -1]$: array of indices from $\set{\mathbf{none},0,1,\ldots,N-1}$\newline $\mathit{header}[0\twodots \lg
    \bar{S}]$: array of offsets, $\mathit{header}[h] = 2^h -1 \mbox{ for
    } h\in \set{0,1,\ldots, \lg \bar{S}}$}

$\mathit{current} \leftarrow \mathit{bucket}\mbox{\rm
  -}\mathit{start} / \ceils{N/\bar{S}}$\;
$\mathit{leaf} \leftarrow \mathit{header}[\lg \bar{S}] + \mathit{current}$\;
$T[\mathit{leaf}] \leftarrow \mathit{bucket}\mbox{\rm -}\mathit{min}$\;
\For {$\ell \in \set{\lg \bar{S}, \lg \bar{S} - 1,\ldots, 1}$} {
  $\mathit{parent} \leftarrow \floors{\mathit{current} / 2}$\;
  $\mathit{this} \leftarrow
    \mathit{header}[\ell] + \mathit{current}$\;
  $\mathit{sibling} \leftarrow
  \mbox{ \textbf{if} } \mathit{current} \bmod 2 = 0\mbox{\,: } 
  \mathit{this} + 1 \mbox{ \textbf{else} } \mathit{this} - 1$\;
  $\mathit{above} \leftarrow
    \mathit{header}[\ell - 1] + \mathit{parent}$\;
  \If {$T[\mathit{sibling}] = \mathbf{none}$} {
    $T[\mathit{above}] \leftarrow
    T[\mathit{this}]$\;
  }
  \ElseIf {$T[\mathit{this}] = \mathbf{none}$} {
    $T[\mathit{above}] \leftarrow
    T[\mathit{sibling}]$\; 
  }
  \ElseIf {$A[T[\mathit{this}]] < 
A[T[\mathit{sibling}]]$} {
    $T[\mathit{above}] \leftarrow
    T[\mathit{this}]$\;  
  }
  \Else {
    $T[\mathit{above}] \leftarrow
    T[\mathit{sibling}]$\; 
  }
  $\mathit{current} \leftarrow \mathit{parent}$
}
\end{minipage}
\caption{Submersion when done at one go.}
\label{fig:submersion}
\end{algorithm2e}

\begin{algorithm2e}[tb!]
\SetStartEndCondition{ }{}{}%
\SetKwInput{Proc}{procedure}
\SetKwInput{Input}{input}
\SetKwInput{Assert}{assert}
\SetKwInput{Data}{data}  
\SetKwIF{If}{ElseIf}{Else}{if}{{\rm :}}{else if}{else{\rm :}}{}%
\DontPrintSemicolon
\Proc{$\mathit{insert}$}
\Input{$i$: index of an element to be inserted into $T$}
\Data{$N$: number of elements; $\bar{S}$:
  workspace target rounded to a power of 2\newline
$A[0\twodots N-1]$: read-only array of elements\newline $T[0\twodots 2\bar{S}
    -1]$: array of indices from $\set{\mathbf{none},0,1,\ldots,N-1}$\newline 
$\mathit{insertion}\mbox{\rm -}\mathit{buffer}\mbox{\rm
    -}\mathit{start}$: index of the beginning of the insertion buffer\newline
$\mathit{insertion}\mbox{\rm -}\mathit{buffer}\mbox{\rm -}\mathit{size}$: number of elements in the
  insertion buffer\newline
$\mathit{insertion}\mbox{\rm -}\mathit{buffer}\mbox{\rm
    -}\mathit{min}$: index of the minimum of the
  insertion buffer}

\Assert{$\mathit{insertion}\mbox{\rm -}\mathit{buffer}\mbox{\rm
    -}\mathit{start} + \mathit{insertion}\mbox{\rm
    -}\mathit{buffer}\mbox{\rm -}\mathit{size} = i$}

$\mathit{insertion}\mbox{\rm -}\mathit{buffer}\mbox{\rm
  -}\mathit{size} \leftarrow \mathit{insertion}\mbox{\rm -}\mathit{buffer}\mbox{\rm
  -}\mathit{size} + 1$\;
\If {$\mathit{insertion}\mbox{\rm -}\mathit{buffer}\mbox{\rm -}\mathit{min} = \mathbf{none} \mbox{
    \rm\textbf{or} } A[i] < A[\mathit{insertion}\mbox{\rm -}\mathit{buffer}\mbox{\rm -}\mathit{min}]$} {
  $\mathit{insertion}\mbox{\rm -}\mathit{buffer}\mbox{\rm -}\mathit{min} \leftarrow i$\;
}
\If {$\mathit{overall}\mbox{\rm -}\mathit{min} = \mathbf{none} \mbox{
    \rm\textbf{or} } A[i] < A[\mathit{overall}\mbox{\rm -}\mathit{min}]$} {
  $\mathit{overall}\mbox{\rm -}\mathit{min} \leftarrow i$\;
}

\If {$\mathit{insertion}\mbox{\rm -}\mathit{buffer}\mbox{\rm -}\mathit{size} = \ceils{N / \bar{S}}$} {
  $\mathit{submersion}\mbox{\rm -}\mathit{buffer}\mbox{\rm -}\mathit{start} \leftarrow
  \mathit{insertion}\mbox{\rm -}\mathit{buffer}\mbox{\rm -}\mathit{start}$\;
  $\mathit{submersion}\mbox{\rm -}\mathit{buffer}\mbox{\rm -}\mathit{size} \leftarrow
  \mathit{insertion}\mbox{\rm -}\mathit{buffer}\mbox{\rm -}\mathit{size}$\;
  $\mathit{submersion}\mbox{\rm -}\mathit{buffer}\mbox{\rm -}\mathit{min} \leftarrow 
\mathit{insertion}\mbox{\rm -}\mathit{buffer}\mbox{\rm -}\mathit{min}$\;
  $\mathit{insertion}\mbox{\rm -}\mathit{buffer}\mbox{\rm -}\mathit{start} \leftarrow i+1$\;
  $\mathit{insertion}\mbox{\rm -}\mathit{buffer}\mbox{\rm -}\mathit{size} \leftarrow 0$\;
  $\mathit{insertion}\mbox{\rm -}\mathit{buffer}\mbox{\rm -}\mathit{min} \leftarrow \mathbf{none}$\;
}
execute $O(1)$ steps of the submersion process, if one is in progress

\caption{Inserting an element into an adjustable tournament tree.}
\label{fig:insert}
\end{algorithm2e}

The pseudo-code of \Insert{} is given in Figure~\ref{fig:insert}.  At
first, the next element from the input array becomes part of the
insertion buffer. If the new element is smaller than the buffer
minimum and/or the overall minimum, the positions of these minima are
updated.  Once the insertion buffer becomes full, the submersion
buffer must have been already submerged into the tournament tree.  At
this point, we treat the insertion buffer as the new submersion buffer
and start a new incremental submersion process that recomputes the
indices at the nodes on the updating path bottom-up, one by one.  As
long as the submersion is not finished, each \Insert{} carries out a
constant amount of the submersion work. Since the work needed to
update this path is $O(\lg S)$, which is $O(N/S)$ when $S \leq N/\lg N$, the process
terminates before the insertion buffer becomes again full.  
Clearly, \Insert{} takes $O(1)$ worst-case time.

\begin{algorithm2e}[tb!]
\begin{minipage}{1.1\textwidth}
\SetStartEndCondition{ }{}{}%
\SetKwInput{Proc}{procedure}
\SetKwInput{Input}{input}
\SetKwInput{Assert}{assert}
\SetKwInput{Data}{data}  
\SetKwFor{For}{for}{{\rm :}}{}
\SetKwIF{If}{ElseIf}{Else}{if}{{\rm :}}{else if}{else{\rm :}}{}%
\DontPrintSemicolon
\Proc{$\mathit{extract}$}
\Input{$j$: index of an element to be extracted from $T$}
\Data{$N$: number of elements; $\bar{S}$:
  workspace target rounded to a power of 2\newline
$A[0\twodots N-1]$: read-only array of elements\newline $T[0\twodots 2\bar{S}
    -1]$: array of indices from $\set{\mathbf{none},0,1,\ldots,N-1}$\newline 
$\mathit{insertion}\mbox{\rm -}\mathit{buffer}\mbox{\rm
    -}\mathit{start}$: index of the beginning of the insertion buffer\newline
$\mathit{insertion}\mbox{\rm -}\mathit{buffer}\mbox{\rm -}\mathit{min}$: index of the minimum
of the
  insertion buffer\newline
$\mathit{submersion}\mbox{\rm -}\mathit{buffer}\mbox{\rm
    -}\mathit{start}$: index of the beginning of the submersion buffer\newline
$\mathit{submersion}\mbox{\rm -}\mathit{buffer}\mbox{\rm -}\mathit{min}$: index of the minimum
of the
  submersion buffer}

$\mathit{latest}\mbox{\rm -}\mathit{output} \leftarrow A[j]$\;
$\mathit{bucket}\mbox{\rm -}\mathit{start} \leftarrow \ceils{N/\bar{S}} \cdot \floors{j / \ceils{N/\bar{S}}}$\;
$\mathit{bucket}\mbox{\rm -}\mathit{min} \leftarrow \mathbf{none}$\;
\For {$i \in \set{\mathit{bucket}\mbox{\rm -}\mathit{start}, \mathit{bucket}\mbox{\rm -}\mathit{start} + 1, \ldots,\min\!\set{\mathit{bucket}\mbox{\rm -}\mathit{start} + \ceils{N/\bar{S}} - 1,
      N - 1}}$} {
  \If {$\mathit{latest}\mbox{\rm -}\mathit{output} < A[i] \mbox{
      \rm\textbf{and} } \left(\mathit{bucket}\mbox{\rm -}\mathit{min} = \mathbf{none}
       \mbox{ \rm\textbf{or} } A[i] < A[\mathit{bucket}\mbox{\rm
           -}\mathit{min}]\right)$} {
       $\mathit{bucket}\mbox{\rm -}\mathit{min} \leftarrow i$\;
  }
}
\If {$\mathit{bucket}\mbox{\rm -}\mathit{start} = \mathit{insertion}\mbox{\rm
    -}\mathit{buffer}\mbox{\rm -}\mathit{start}$} {
  $\mathit{insertion}\mbox{\rm -}\mathit{buffer}\mbox{\rm
    -}\mathit{min} \leftarrow \mathit{bucket}\mbox{\rm
    -}\mathit{min}$\;
}

\ElseIf {$\mathit{bucket}\mbox{\rm -}\mathit{start} = \mathit{submersion}\mbox{\rm
    -}\mathit{buffer}\mbox{\rm -}\mathit{start}$} {
  $\mathit{submersion}\mbox{\rm -}\mathit{buffer}\mbox{\rm
    -}\mathit{min} \leftarrow \mathit{bucket}\mbox{\rm
    -}\mathit{min}$\;
  $\mathit{submersion}(\mathit{submersion}\mbox{\rm
    -}\mathit{buffer}\mbox{\rm -}\mathit{start}, \mathit{submersion}\mbox{\rm
    -}\mathit{buffer}\mbox{\rm -}\mathit{min})$\;
$\mathit{submersion}\mbox{\rm
    -}\mathit{buffer}\mbox{\rm -}\mathit{min} \leftarrow
  \mathbf{none}$\;
}
\Else {
$\mathit{submersion}(\mathit{bucket}\mbox{\rm -}\mathit{start}, \mathit{bucket}\mbox{\rm -}\mathit{min})$
}
$\mathit{overall}\mbox{\rm -}\mathit{min} \leftarrow \mathbf{none}$\;
\For {$k \in \set{T[0], 
\mathit{submersion}\mbox{\rm -}\mathit{buffer}\mbox{\rm
  -}\mathit{min},
\mathit{insertion}\mbox{\rm -}\mathit{buffer}\mbox{\rm
  -}\mathit{min}}$} {
  \If {$k \neq \mathbf{none} \mbox{ \rm\textbf{and} } \left(\mathit{overall}\mbox{\rm -}\mathit{min} = \mathbf{none}
      \mbox{ \rm\textbf{or} } A[k] < A[\mathit{overall}\mbox{\rm
          -}\mathit{min}]\right)$} {
      $\mathit{overall}\mbox{\rm -}\mathit{min} \leftarrow k$\;
  }
}
\end{minipage}

\caption{Extracting an element from an adjustable tournament tree.}
\label{fig:extract}
\end{algorithm2e}

In \Extract{}, there are three cases depending on whether the bucket
that contains the element to be extracted is one of the buffers or
is covered by a node of the tournament tree.  However, these cases are quite
similar. For a pseudo-code description, see Figure~\ref{fig:extract}.
To begin with, the latest output is set up to date.  The bucket index
of the given element can be determined by simple calculations.
Because of the monotonicity assumption the smallest alive element 
is to be extracted, so the bucket must be scanned to find
its new minimum.  If the current bucket is the insertion buffer, it is
just enough to update the position of its minimum.  If the current
bucket is the submersion buffer, the submersion process is completed
by recomputing the indices at the nodes on the updating path covering the
submersion buffer. Hereafter the submersion buffer ceases to exist.
If the current bucket is covered by the tournament tree, it is
necessary to recompute the indices at nodes on the updating path covering
the current bucket. At the end, the position of the overall minimum is
to be updated.  It is the scanning of a bucket that makes this
operation expensive: The worst-case running time is $O(N/S + \lg S)$,
which is $O(N/S)$ when $S \leq N/\lg N$.

\begin{figure}[tb!]
\begin{center}
\centerline{\includegraphics[scale=0.41]{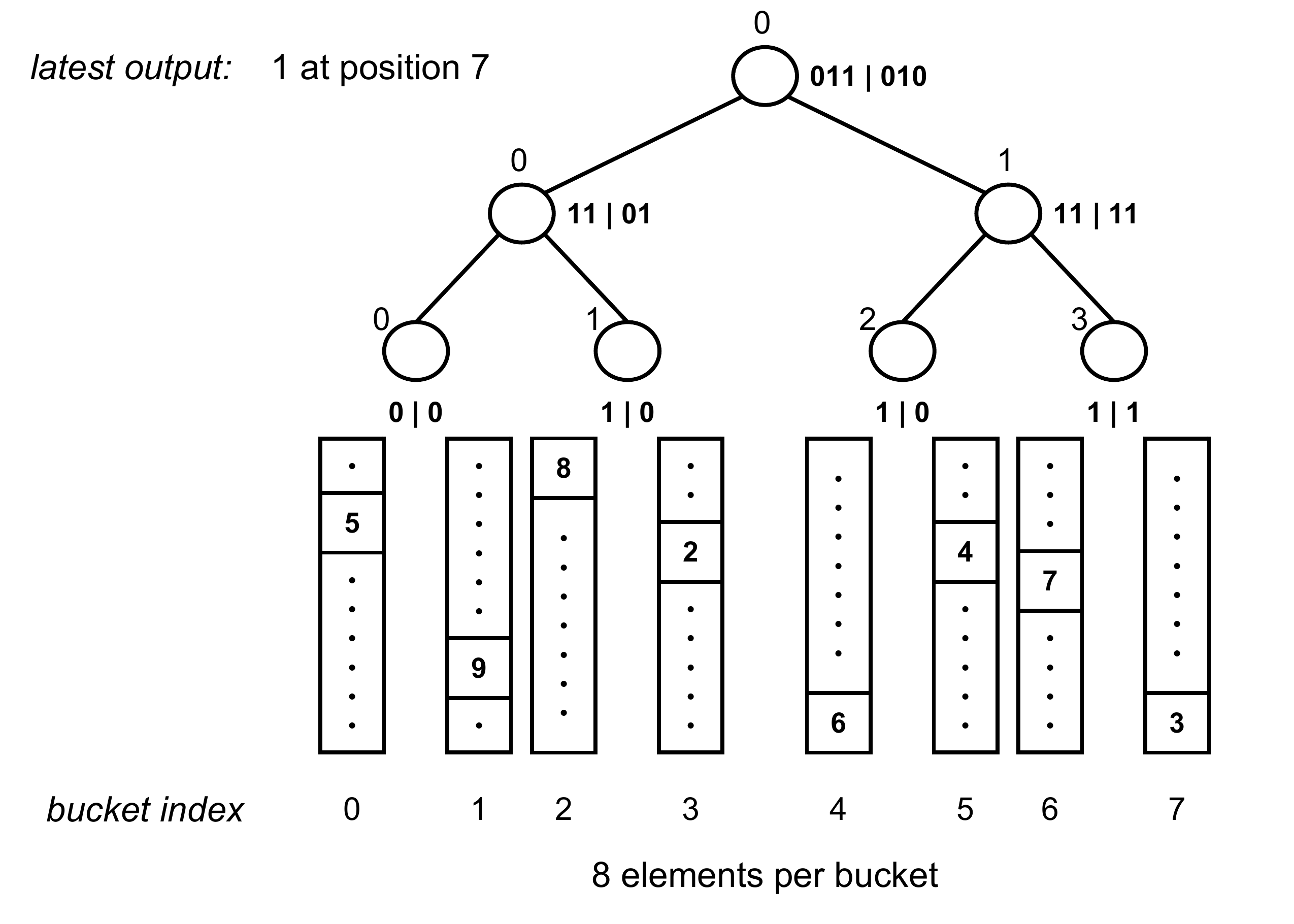}}
\end{center}

\vspace*{-0.5cm}
\caption{An adjustable navigation pile corresponding to the adjustable
  tournament tree in \mbox{Figure}~\ref{fig:tournament}. In this snapshot all
  buckets contain one or more alive elements so the bit vector---that
  is not shown---indicating their existence contains just 1 bits.}
\label{fig:pile}
\end{figure}

\subsection{Adjustable Navigation Piles} 
In brief, an \emph{adjustable navigation pile} is a compact
representation of an adjustable tournament tree.  The main differences
are as follows (compare Figure~\ref{fig:tournament} and
Figure~\ref{fig:pile}):
\begin{enumerate}
\item A bit vector of size $2\bar{S} -1$ is used to indicate
  whether or not the buckets covered by each node contain any alive
  elements. As for the data in the tournament tree, in this bit vector the
  bits of the nodes are stored in breadth-first order. This bit vector
  can be used to emulate the constant $\mathbf{none}$.
\item Only branch nodes, i.e.~nodes whose heights are larger than
  0, store some additional information about the position of the
  smallest alive element in the covered range of each of these nodes. In the
  complete binary tree built above the buckets, the number of branch nodes
  is $\bar{S} - 1$.
\item The navigation information is stored in a bit vector of
  size $4\bar{S}$ (an explanation will follow shortly).  To save space,
  at the bottom of the tree, the position of the smallest alive
  element is only specified approximately. Here the details of our
  construction differ from those used in the navigation piles of
  \cite{KV03} and their precursors \cite{PR98}, although the
  techniques used are similar.
\end{enumerate}

A branch node of height $h \in \set{1,2,\ldots,\lg \bar{S}}$ covers
$2^h$ buckets. As in a navigation pile \cite{KV03}, due to scarcity of bits, 
the bucket index is \emph{relative} within the covered range. We use $h$ bits to
specify in which bucket the smallest alive element is.  
A significant new ingredient, borrowed from \cite{PR98}, is the concept of a
\emph{quantile}. For a branch node of height
$h$, every covered bucket is divided into $2^h$ quantiles, and we
store additional $h$ bits to specify in which quantile the smallest
alive element is.  We call this quantile the {\it active quantile} of the node. 
A quantile contains $\ceils{N/(\bar{S} \cdot 2^h)}$
elements, except that the last quantile can possibly be smaller.
Thus, we use $2h$ bits per node; but if $2h \geq \ceils{\lg{N}}$, we
only use $\ceils{\lg{N}}$ bits (since this is enough to
specify the exact position of the smallest alive element).  To sum up,
since there are $\bar{S}/2^h$ nodes of height $h$ and since at each
node we store $\min\!\set{2h, \ceils{\lg{N}}}$ bits, the total number of
navigation bits is bounded by
$$\sum_{h=1}^{\lg \bar{S}} \frac{\bar{S} \cdot \min\!\set{2h,
    \ceils{\lg{N}}}}{2^h} < 4\bar{S}\,.$$

The navigation bits are stored in a bit vector in breadth-first order.
As before, we maintain a \emph{header} giving the position of the first bit
at each level. The space needed by the header is $O(\lg^2 \bar{S})$
bits. Inside each level, the navigation information is stored compactly
side by side, and the nodes are numbered at each level starting from
$0$.  Since the length of the navigation bits is fixed for all nodes
at the same level, using the height and the index of a node, it is
easy to calculate the positions where the navigation bits of that node
are stored.

Let us now consider how to access the desired quantile for a branch
node. The procedure is summarized in pseudo-code form
in Figure~\ref{fig:quantile}. Let the branch node be at index $v$ within
its own level, and assume that its height is $h$. The first element of
the covered range of this branch node is in position $v \cdot 2^h \cdot
\ceils{N/\bar{S}}$.  The first $h$ bits of the navigation information
give the desired bucket inside the covered range; let the relative
bucket index be $b$, so we have to
go $b \cdot \ceils{N/\bar{S}}$ positions forward. The second $h$ bits
of the navigation information give the desired quantile inside that
bucket; let this quantile be $q$, so we have to proceed another
$q \cdot \ceils{N/(\bar{S} \cdot 2^h)}$ positions forward before we
reach the beginning of the desired quantile. Obviously, these
calculations can be carried out in constant time.

\begin{algorithm2e}[tb!]
\SetStartEndCondition{ }{}{}%
\SetKwInput{Proc}{procedure}
\SetKwInput{Input}{input}
\SetKwInput{Assert}{assert}
\SetKwInput{Data}{data}  
\SetKwFor{For}{for}{{\rm :}}{}
\SetKwIF{If}{ElseIf}{Else}{if}{{\rm :}}{else if}{else{\rm :}}{}%
\DontPrintSemicolon
\Proc{$\mathit{calculate}\mbox{\rm -}\mathit{quantile}$}
\Input{$v$: index of a branch node at its own level; $h$: height of that node}
\Data{$N$: number of elements; $\bar{S}$:
  workspace target rounded to a power of 2\newline
 $B[0\twodots 4\bar{S}-1]$: array of navigation bits\newline $\mathit{header}[1\twodots \lg
    \bar{S}]$: array of offsets}

$\lambda \leftarrow \min\!\set{2h, \ceils{\lg N}}$\;
$\mathit{info}\mbox{\rm
  -}\mathit{start} \leftarrow \mathit{header}[h] + v \cdot \lambda$\; 
\If {$\lambda = \ceils{\lg N}$} {
  $\mathit{quantile}\mbox{\rm -}\mathit{start} \leftarrow
  B[\mathit{info}\mbox{\rm -}\mathit{start}\twodots
    \mathit{info}\mbox{\rm -}\mathit{start} + \ceils{\lg N} - 1]$\;
  $\mathit{quantile}\mbox{\rm -}\mathit{size} \leftarrow 1$\;
   \textbf{return} $(\mathit{quantile}\mbox{\rm -}\mathit{start}, \mathit{quantile}\mbox{\rm -}\mathit{size})$\;
}

$b \leftarrow
  B[\mathit{info}\mbox{\rm -}\mathit{start}\twodots
    \mathit{info}\mbox{\rm -}\mathit{start} + h - 1]$\;

$q \leftarrow
  B[\mathit{info}\mbox{\rm -}\mathit{start} + h\twodots
    \mathit{info}\mbox{\rm -}\mathit{start} + 2\cdot h - 1]$\;

$\mathit{covered}\mbox{\rm -}\mathit{range}\mbox{\rm -}\mathit{start}
\leftarrow v \cdot 2^h \cdot \ceils{N/\bar{S}}$\;

$\mathit{bucket}\mbox{\rm -}\mathit{start}
\leftarrow 
\mathit{covered}\mbox{\rm -}\mathit{range}\mbox{\rm -}\mathit{start} +
b \cdot \ceils{N/\bar{S}}$\;

$\mathit{quantile}\mbox{\rm -}\mathit{start}
\leftarrow 
\mathit{bucket}\mbox{\rm -}\mathit{start} + q \cdot
\ceils{N/(\bar{S}\cdot 2^h)}$\;

$\mathit{quantile}\mbox{\rm -}\mathit{past}
\leftarrow 
\min\!\set{\mathit{quantile}\mbox{\rm -}\mathit{start} +
  \ceils{N/(\bar{S}\cdot 2^h)}, \mathit{bucket}\mbox{\rm
    -}\mathit{start} + \ceils{N/\bar{S}}}$\;

$\mathit{quantile}\mbox{\rm -}\mathit{size}
\leftarrow 
\mathit{quantile}\mbox{\rm -}\mathit{past} -
\mathit{quantile}\mbox{\rm -}\mathit{start}$\;

   \textbf{return} $(\mathit{quantile}\mbox{\rm -}\mathit{start}, \mathit{quantile}\mbox{\rm -}\mathit{size})$\;

\caption{Calculating the beginning and size of the active quantile of a node.}
\label{fig:quantile}	
\end{algorithm2e}

The priority-queue operations are implemented in a similar way as
for an adjustable tournament tree. To facilitate constant-time
\Minimum{}, we keep in memory the index of the overall minimum (since
the root of an adjustable navigation pile does not necessarily specify
a single element).  In \Insert{} and \Extract{}, the subtle difference
is dealing with quantiles. When updating the
navigation information for a branch node, at the bottom of an
adjustable navigation pile, we do not have direct access to the
minimum among the alive elements covered. Instead, we have to scan the
elements in the quantiles specified for the sibling nodes of the nodes
along the updating path.  Later on, a quantile is said to be
\emph{active} if it contains the minimum among the alive elements
covered by a node.  After updating the navigation bits of a node $v$,
we locate its parent node $u$ and its sibling node $w$. The navigation bits of
$w$ are used to locate its active quantile.  This quantile is
scanned, and the minimum of the alive elements is found and compared
with the minimum of the alive elements covered by $v$. From the bucket
index and the position of the smaller of the two elements, the
navigation bits of $u$ are then calculated and accordingly
updated. If the active quantile of $u$ has only one element, the
position of this single element can be stored as such.

The key point is that for a node of height $h$ the size of the active
quantile is at most $\ceils{N/(\bar{S} \cdot 2^h)}$, so the total work
done in the scans of the quantiles of the siblings along the updating
path is proportional to $\sum_{h=1}^{\lg \bar{S}} \ceils{N/(\bar{S}
  \cdot 2^h)}$, which is $O(N/S + \lg S)$.  It follows that the
asymptotic efficiency of the priority-queue operations is the same as for an
adjustable tournament tree.

\subsection{Getting Rid of the Assumptions}  So far we have consciously
ignored the fact that the sizes of the buckets depend on $N$, 
and that we might not know this value beforehand. The
standard way of handling this is to rely on global
rebuilding \cite[Chapter V]{Ove83}. We use an estimate $N_0$ and initially set $N_0 = 8$.
We build two data structures, one for $N_0$ and another for $2N_0$. The first structure
is used to perform the priority-queue operations, but insertions and
extractions  are mirrored in the second structure (if the extracted element exists there).  
When the structure for $N_0$ becomes too small, we dismiss the smaller structure
in use, double $N_0$, and in accordance start building a new structure of size $2N_0$. 
We should speed up the construction of the new structure by inserting up to two
alive elements into it at a time, instead of only one. This guarantees
that the new structure will be ready for use before the first one
is dismissed.  Even though global rebuilding makes the
construction more complicated, the time and space
bounds remain asymptotically the same.

Since we have random-access capability to the read-only input, it is
not necessary that elements are inserted by visiting the input
sequentially, but insertions should still be monotonic. If
this is the case, in connection with each \Insert{}, we have to fix
the information related to the current bucket as in \Extract{}. That
is, we have to find the smallest alive element of the bucket and, if
the inserted element is smaller than the current minimum at that
bucket, update the navigation information on the updating path covering
that bucket. The worst-case cost of \Insert{} then becomes
the same as that of \Extract{}, i.e.~$O(N/S + \lg S)$.

In some applications, insertions and extractions may 
not be monotonic. To handle this situation, we have to allocate one bit per array entry
($N$ bits in total), indicating whether the corresponding element is alive or not.  

\section{Sorting}
\label{sec:sorting}

\subsection{The Pagter-Rauhe Algorithm}
Let us turn our attention to sorting. Given $N$ elements in
a read-only array, the task is to print the elements in
non-decreasing order. Assume that the asymptotic workspace target is $S$.
In the basic setting, Pagter and Rauhe \cite{PR98} proved that
the running time of their sorting algorithm is $O(N^2/S + N \lg^2 S)$
using $O(S)$ bits of workspace. Lagging behind the optimal bound for
the space-time product by a logarithmic factor when $S=\omega(N/\lg^2
N)$, they suggested using their memory-adjustable data structure in
Frederickson's adjustable binary heap \cite{Fre87} to handle
subproblems of size $(N\lg N)/S$ using $O(\lg N)$ bits for each.  In
accordance, by combining the two data structures, they
achieve an optimal $O(N^2/S)$ running time for sorting
when $\lg N \leq S \leq N/\lg N$.  In our treatment we avoid the
complication of plugging two data structures together.

\subsection{Priority-Queue Sort} 
To sort the elements, we create an empty adjustable navigation
pile, insert the elements into this pile by scanning the read-only
array from beginning to end, and then repeatedly extract the minimum of the
remaining elements from the pile. The pseudo-code in
Figure~\ref{sorting-algorithm} implements this algorithm.

\begin{algorithm2e}[tb!]
\SetStartEndCondition{ }{}{}%
\SetKwInput{Proc}{procedure}
\SetKwInput{Input}{input}  
\SetKwFor{For}{for}{{\rm :}}{}
\SetKwFor{While}{while}{{\rm :}}{}
\SetKwIF{If}{ElseIf}{Else}{if}{{\rm :}}{else if}{else{\rm :}}{}%
\DontPrintSemicolon
\Proc{$\mathit{priority}${\rm -}$\mathit{queue}${\rm -}$\mathit{sort}$}
\Input{$A[0\twodots N-1]$: read-only array of $N$ elements; $S$: workspace target}

$P \leftarrow \mathit{navigation}${\rm -}$\mathit{pile}(A, S)$\;
\For {$i \in \set{0,1,\ldots, N-1}$} {
  $P.\Insert{}(i)$\;
}
\While {$|P| > 0$} {
  $j \leftarrow P.\Minimum{}()$\;
  $P.\Extract{}(j)$\;
  $\mathit{print}(A[j])$\;
}
\caption{Priority-queue sort; the position of an
  element is specified by its index.}
\label{sorting-algorithm}	
\end{algorithm2e}

\subsection{Analysis}  From the bounds derived for the priority queue,
the asymptotic performance can be directly deduced: The worst-case
running time is $O(N^2/S + N \lg S)$ and the size of workspace is
$O(S)$ bits, where $S \geq \lg N$. It is also easy to count the number of element
comparisons performed during the execution of the algorithm. When
inserting the $N$ elements into the data structure, $O(N)$ element
comparisons are performed. We can assume that after these insertions,
the buffers are submerged into the main structure. In each \Extract{}
we have to find the minimum of a single bucket which requires at most
$N/\bar{S}$ element comparisons. In addition, we have to
update a single path in the complete binary tree. At each level, the
minimum below the current node is already known and we have to scan the
quantiles of the sibling nodes. During the path update, we have to perform
at most $N/\bar{S} + \lg \bar{S}$ element comparisons.
Hence, the total number of element comparisons performed is bounded by $2N^2/\bar{S} + N \lg \bar{S} + O(N)$, 
which is $2N^2/S + N \lg S + O(N)$ since $S \leq \bar{S} \leq 2S$. 

\section{Augmenting the Adjustable Navigation Pile}
\label{aug}

\subsection{Motivation for Augmentation} 
\begin{figure}[t!]
\centerline{\includegraphics[scale=0.41]{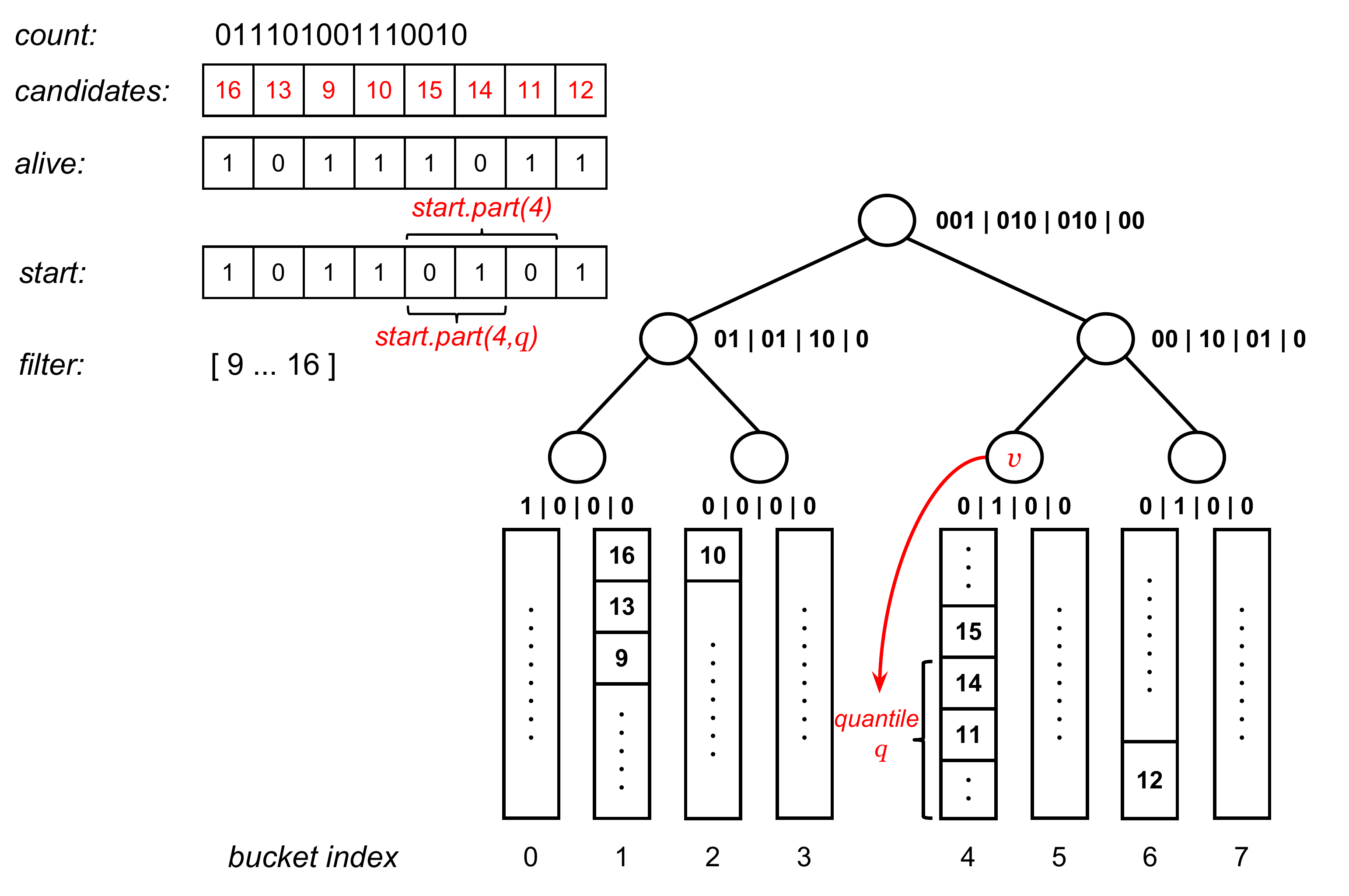}}
\caption{A snapshot of an augmented navigation pile that has $N = 64$
  and $S = 8$. Only elements within the range of the filters are
  shown. The snapshot is taken considering the given $\mathit{alive}$
  vector. For each node, the navigation bits are displayed in
  left-to-right order as follows: relative bucket index, quantile
  index, $\mathit{start.part}$ index, and $\mathit{start.before}$. As
  an example, we can see that node $v$ refers to the candidate $11$ as
  its minimum alive element. Hence, its relative bucket index is $0$,
  which refers to bucket $4$. Bucket $4$ is partitioned into two
  quantiles in the view of node $v$, where the candidate $11$ lies in
  the second quantile $q$. So, the quantile index for node $v$ is
  $1$. $\mathit{start.part}(4)$ in $\mathit{start}$ is dedicated to
  bucket $4$. At node $v$, $\mathit{start.part}(4)$ is partitioned
  into two parts.  Since the candidate $14$ is 
  $\FirstCandidate(q)$, and as this candidate is in the
  first part of $\mathit{start.part}(4)$, the start.part index of node
  $v$ is set to $0$.  The part that contains $\FirstCandidate(q)$
  is referred to as $\mathit{start.part}(4,q)$.}	
\label{ANP}
\vspace{-0.5cm}
\end{figure}

In our algorithm for computing the convex hull of a set of planar points (see Section~\ref{ch}), 
insertions are neither monotonic nor sequential and extractions are not monotonic. 
Nevertheless, we want to keep the algorithm memory adjustable and use $O(S)$ bits of extra
space for $S < N$.  To limit the size of working space, our
solution is to work with a subset of the input constituting at most
$S$ elements at a time. In more details, the input is processed in a
number of rounds, where in each round we shall have two filter
elements and only elements whose values are between these filters are
to be inserted or extracted from the adjustable navigation pile. We
refer to these $S$ elements as the \emph{candidates}.  Note that
the candidates need not be contiguous in the read-only input array.  
In every round, we use a vector of $S$ bits, one bit per candidate, 
to indicate whether each of these candidates is still alive or has been deleted.
Subsequently, we need to map the indices of the input array to
indices in the range $[0 \twodots S-1]$. To be able to do that,
the adjustable navigation pile needs to be augmented with additional
information. We shall refer to the adjustable navigation pile explained in
Section \ref{sec:priority-queues} as the \emph{unaugmented navigation
  pile} to distinguish it from the augmented version to be described
in this section.  Next, we introduce the additional structures used in
the augmentation. Then, we explain how to update and utilize these
structures in the priority-queue operations.  A schematic view
illustrating the components of an augmented navigation pile is given in
Figure~\ref{ANP}.

\subsection{Additional Structures} 
For the sake of simplicity, assume that the required workspace target $S$ is
a power of 2.  We augment the adjustable navigation pile with the
following data structures:
\begin{itemize}
\item A bit vector $\mathit{alive}$ of size $S$ is used to denote
  whether each candidate is currently alive or not. The order of the
  candidates in this vector is identical to their order in the read-only
  array. The $\mathit{alive}$ vector is dynamically updated by \Insert{}
  and \Extract{} operations.

\item A bit vector $\mathit{start}$ of size $S$---corresponding to the same
  elements, in the same order, as $\mathit{alive}$---is
  used to denote whether a candidate is the first, among other
  candidates, of an active quantile or not. We refer to the candidate corresponding to the first entry for a  
  candidate from quantile $q$ as $\FirstCandidate(q)$.
	(Note that a candidate may simultaneously be the first candidate of more than one active quantile.) 
	The $\mathit{start}$ vector is also dynamic, as every
  \Insert{} and \Extract{} may change it.  Every bucket will possibly map
  to a part of $\mathit{alive}$ and $\mathit{start}$, which obviously has at
  most $\lceil N/S \rceil$ entries.  We refer to the part of $\mathit{start}$
  corresponding to bucket $u$ as $\mathit{start.part}(u)$.

\item A static bit vector $\mathit{count}$ is used to store the number
  of candidates contained in each bucket. We encode these counts in
  unary, using a $0$ bit to mark the border between every two
  consecutive buckets. Since we are dealing with at most $S$
  candidates, the vector contains at most $S$ ones; and since we have
  exactly $S$ buckets, it contains $S-1$ zeros. The $\mathit{count}$
  vector should efficiently support \Rank{} and \Select{} queries. The
  bit vector and the accompanying rank-select structures thus consume
  $\Theta(S)$ bits.  The objective is to efficiently locate the first
  entry of any bucket in $\mathit{alive}$ and $\mathit{start}$ using the index of the bucket. 
	Assume the first bucket has index $0$. Let $u > 0$ be the index of the bucket whose first
  entry in $\mathit{alive}$ and $\mathit{start}$ is to be located. Then, $t =
  \mathit{count}$.\Select$0(u)$ is the index of the last $0$ bit
  preceding bucket $u$ in $\mathit{count}$.  It follows that $t-u+1$
  is the number of candidates lying in the buckets $0$, $1$, \dots,
  $u-1$, which precede the first entry of bucket $u$ in
  $\mathit{alive}$ and $\mathit{start}$. The size (number of bits)
  of $\mathit{start.part}(u)$ is calculated as 
	$\mathit{count}$.\Select$0(u+1)$ $-$ $\mathit{count}$.\Select$0(u) - 1$.

\item For every node of height $h$, associated with bucket $u$ and
  quantile $q$, $\mathit{start.part}(u)$ is further divided into $2^h$
  subparts (or less, if its size is less than $2^h$). Another $h$ bits
  will be stored in every node of height $h$ to indicate in which
  subpart $\FirstCandidate(q)$ lies. We refer to
  this subpart as $\mathit{start.part}(u,q)$. Since the size of
  $\mathit{start.part}(u)$ is at most $\lceil N/S \rceil$, the size of
  $\mathit{start.part}(u,q)$ is bounded from above by $\ceils{N/(S
    \cdot 2^h)}$, which is the size of a quantile.

\item For every branch node of height $h$, associated with bucket $u$ and
  quantile $q$, an additional $\left\lceil{\lg h}\right\rceil$ bits
  will be used. These bits encode the number of first candidates
  of other active quantiles that exist in the same subpart
  $\mathit{start.part}(u,q)$ before $\FirstCandidate(q)$, 
  i.e.~the number of ones in $\mathit{start.part}(u,q)$
  preceding the one representing $\FirstCandidate(q)$. 
  Let us refer to the value of this number as $\mathit{start.before}(u,q)$.  
	The reason we need only $\left\lceil{\lg h}\right\rceil$ bits to store this
  information for each node of height $h$ is that only quantiles tied
  to nodes (whose heights are less than $h$) on one and only one path
  from a leaf node to the node associate with $q$ can have their first candidates
  before $\FirstCandidate(q)$ in $\mathit{start.part}(u,q)$.
\end{itemize}

It directly follows, using simple calculations, that the space
complexity of the augmented navigation pile is still $\Theta(S)$ bits. 

To keep the time complexity for \Insert{} and \Extract{} in $O(N/S +
\lg S)$, we need to know, in an efficient way, if a given candidate that
belongs to an active quantile is alive or not.  Starting with the
index of a candidate in the read-only array, we want to find, without
altering the time bounds, the index of the corresponding bit in $\mathit{alive}$.
Given a node $v$, associated with bucket $u$ and quantile $q$,
the index of $\FirstCandidate(q)$ in
$\mathit{alive}$ is to be located. Using the $\mathit{count}$ vector
and the navigation bits of node $v$, we can easily locate
$\mathit{start.part}(u,q)$, where the entry we are searching for
lies. We also get the value $r = \mathit{start.before}(u,q)$ from the
navigation bits of node $v$. As previously stated, the size of
$\mathit{start.part}(u,q)$ is at most the size of quantile $q$. While
visiting node $v$, we shall be scanning quantile $q$ anyhow. A scan of
$\mathit{start.part}(u,q)$ would then not alter the worst-case
asymptotic time complexity. We scan $\mathit{start.part}(u,q)$ to find
the $r$th one bit, the index of the entry at which we find this bit is
the index of $\FirstCandidate(q)$ in $\mathit{alive}$.

We always access the elements of a quantile
sequentially. To locate the corresponding elements in
$\mathit{alive}$, we start at the first entry of the quantile in
$\mathit{alive}$ as explained above. While
scanning the quantile, we repeatedly check the elements one after
another. If the next element is a candidate (lying in the range of the
filters), we increase the current index to the next entry in
$\mathit{alive}$ to correspond to this candidate.  The same procedure
can be applied on buckets. We get the first entry of the bucket
in $\mathit{alive}$ using the $\mathit{count}$ vector, and then move
sequentially on the bucket and on $\mathit{alive}$, increasing the
$\mathit{alive}$ index whenever we encounter the next candidate in the bucket.

\subsection{Operations} 
We next explain how \Insert{} and \Extract{} can be performed in our
augmented navigation pile in  $O(N/S + \lg S)$ worst-case time per operation.

We first find the bucket in which the element to be inserted or extracted
lies; this can be done with simple calculations in constant time once
we have the array index of the element. Let the index of this bucket
be $u$. Using the $\mathit{count}$ vector, we get the first entry of
this bucket in $\mathit{alive}$ as explained earlier, and move
sequentially on bucket $u$ and $\mathit{alive}$. We can then get the
indices of the candidates lying in this bucket within
$\mathit{alive}$, and consequently know whether each of these
candidates is currently alive or not. After knowing the index of the
element to be inserted or extracted in $\mathit{alive}$, we should set
the corresponding bit to 1 or 0 respectively. Also, while scanning the bucket, we
would know if this element is the minimum in the bucket or not. If the
minimum alive element in bucket $u$ has changed due to the current operation,
the following updates need to be done.

As in an unaugmented navigation pile, the information in the nodes along the updating path is to be fixed bottom up. 
The update will work as in the unaugmented navigation pile, where we scan the quantiles associated with the nodes lying on or hanging from the updating path. However, here we also want to know whether each of the candidates in these quantiles is alive or not.
Before accessing a quantile, we get its first entry in
$\mathit{alive}$; this can be done as previously described. 
Then, we simultaneously scan both the quantile and the corresponding subpart in $\mathit{alive}$.

Next, we show how to update the $\mathit{start}$ vector.  Suppose that
we are to handle a node $v$, associated with quantile $q$, on the updating path, 
knowing that its child has just been handled.  If the index of 
$\FirstCandidate(q)$ is different from the index of the first 
candidate of each of the two quantiles associated with the two children
of node $v$, we reset the bit for $\FirstCandidate(q)$ 
in $\mathit{start}$ to 0.  Note that we do not reset that bit
to 0 if it is the first entry of the quantile of a child of node $v$
in $\mathit{start}$.  Alternatively, this bit may be temporarily reset
to 0 in the previous step and then again set to 1 within the upcoming
step.  Assume that the child of node $v$ that is associated with the quantile
that has the smaller element of the two children of $v$ associated with quantile
$q'$.  Now the quantile associated with node $v$ should be either the
first half or the second half of quantile $q'$.  If it is the first
half, then the corresponding entry in $\mathit{start}$ must have been
already set to 1 before.  Else, we move sequentially on quantile $q'$
and $\mathit{start}$ till we reach the first candidate in the second
half of quantile $q'$, and set its corresponding entry in $\mathit{start}$ to 1. 
The above procedure is repeated for every node on the updating path.

For the nodes along the updating path, we show next how the additional
bits in our augmented navigation pile will be updated.  Consider a
node $v$ that is associated with bucket $u$ and quantile $q$ after the update.
While handling node $v$, as explained earlier, we are able to know the
entry in $\mathit{start}$ for $\FirstCandidate(q)$ as
well as the size of $\mathit{start.part}(u,q)$.  Knowing these values,
it is easy to update the bits indicating $\mathit{start.part}(u,q)$ in
node $v$. Also, after getting these bits, we loop on
$\mathit{start.part}(u,q)$ to count the number of 1 bits in this subpart
preceding the entry for $\FirstCandidate(q)$ in
$\mathit{start}$, and store this value for node $v$ in $\mathit{start.before}(u,q)$.

It is obvious that the update will be performed on at most $\lg S$ nodes on
the updating path. Also, looping on the quantiles and the
corresponding parts of $\mathit{start}$ for the nodes on the updating
path would sum up to $O(\sum_{i=1}^{\lg S} \lceil N/(S \cdot 2^i)
\rceil )$. So the time complexity for the update is in $O(N/S + \lg
S)$, as claimed.

The following lemma summarizes the functionality of our data structure.

\begin{lemma}
Assume that we are given a read-only array of $N$ elements, and two
filters that enclose at most $S$ of these elements between their
values.  Using $\Theta(S)$ bits of workspace, after spending $O(N)$
worst-case time on building the augmented structure, \Insert{} and
\Extract{} can be applied to any element of the array whose value is
between the filters in $O(N/S + \lg S)$ worst-case time per operation.
\end{lemma}

\section{Convex Hulls}
\label{ch}

\subsection{The Chan-Chen Algorithm}

Consider now the problem of computing the convex hull of a set
of $N$ points in the plane given in a read-only array $P$. Without loss of generality,
we can assume that the points are unique such that no two
points have the same $x$-coordinate or $y$-coordinate. The task is to
print the points on the convex hull in the clockwise order of their
appearance on the hull, starting from some arbitrary point. As is
standard, it is enough to show how to compute the upper hull of the
point set; the lower hull can be computed in a symmetric manner.
We assume the availability of the standard geometric primitive that
tells whether or not there is a right turn on point $P[j]$ when going from
point $P[i]$ to point $P[k]$ via $P[j]$; we denote this predicate as
$\mathit{right}\mbox{\rm -}\mathit{turn}(P[i], P[j], P[k])$.

\begin{algorithm2e}[tb!]
\begin{minipage}{1.0\textwidth}
\SetStartEndCondition{ }{}{}
\SetKwInput{Proc}{procedure}
\SetKwInput{Input}{input}  
\SetKwFor{For}{for}{{\rm :}}{}
\SetKwFor{While}{while}{{\rm :}}{}
\SetKwIF{If}{ElseIf}{Else}{if}{{\rm :}}{else if}{else{\rm :}}{}%
\DontPrintSemicolon
\Proc{$\mathit{compute}${\rm -}$\mathit{convex}${\rm -}$\mathit{hull}$}
\Input{$P[0\twodots N-1]$: read-only array of $N$ points; $s$: workspace target}

$i_0 \leftarrow \mathbf{none}$\;
\For {$i \in \set{0,1,\ldots,N - 1}$} {
  \If {$i_0 = \mathbf{none} \mbox{ \rm\textbf{or} } x$-coordinate$(P[i])$ $<$ $x$-coordinate$(P[i_0])$} {
     $i_0 \leftarrow i$\;
  }
}  

\While {$i_0 \neq \mathbf{none}$} { 

  let $\sigma$ be the vertical slab containing $P[i_0]$ and the next $s-1$ points (if
  possible)\;
~~~~on the right of and closest to the wall determined by $P[i_0]$\; 
  let $\set{i_0,i_1,\ldots, i_{s-1}}$ be the indices of the $s$ points in
  $\sigma$, sorted by $x$-coordinate\; 
  use Graham's scan to compute the upper hull of these points\;
  let $\set{h_0, h_1,h_2,\ldots, h_{s'}}$ be the indices of the points
  on this hull ~~\textbf{//} $h_0 = i_0$\;
  $j' \leftarrow \mathbf{none}$ ~~\textbf{//} right endpoint of the hull edge crossing the right wall of $\sigma$\;
  \For {each point $P[j]$ to the right of the wall determined by $P[i_{s-1}]$} {
    \If {$j' \neq \mathbf{none}$ \rm\textbf{and} $\mathit{right}\mbox{\rm -}\mathit{turn}(P[h_{s'}], P[j'], P[j])$} {
       \textbf{continue}
    }
    \While {$s' > 0 \mbox{ \rm\textbf{and} \rm\textbf{not} } 
\mathit{right}\mbox{\rm -}\mathit{turn}(P[h_{s'-1}], P[h_{s'}], P[j])$} {
      $s' \leftarrow s' - 1$\;
    }
      $j' \leftarrow j$\;
  }
  \For {$k \in \set{h_0,h_1,\ldots,h_{s'}}$} {
    $\mathit{print}(P[k])$\;
  }
  $i_0 \leftarrow j'$\;
}
\end{minipage}

\caption{High-level description of the Chan-Chen algorithm.\label{convex-hull-algorithm}}
\end{algorithm2e}

A high-level description of the algorithm by Chan and Chen \cite{CC07}
is given in Figure~\ref{convex-hull-algorithm}. Our algorithm is
similar but the details are different.  When the working space
of $\Theta(S)$ bits is available, they set the space
parameter $s$ to $S/\lg N$ and use $\Theta(s)$ indices to recall
the points being processed. The algorithm performs $\lceil N/s
\rceil$ rounds and handles the $s$ points with the next smallest 
$x$-coordinates in each round; these points form a vertical slab
$\sigma$. In each round, the algorithm starts with a known hull vertex
$P[i_0]$ and computes the part of the upper hull for the points of
$\sigma$ starting from point $P[i_0]$ and ending at the left endpoint
of the hull edge crossing the right wall of $\sigma$.

When finding the $s$ points with the next smallest $x$-coordinates
among the remaining points, the algorithm uses space for $2s$
indices and maintains in the first half the $s$ indices of the points with the
smallest $x$-coordinates among the points examined so far. Each time,
the second half is refilled with another $s$ indices from the
unexamined portion, the median of the $x$-coordinates of the $2s$
recorded points is found, and these points are partitioned around this
median. This is repeated until all the points are examined, leaving
the $s$ points within the slab $\sigma$. The upper hull of these
points can then be constructed using any
of the known $O(s \lg s)$ convex-hull algorithms \cite[Chapter~3]{PS85}; a natural choice is to use Graham's scan since the rest of
the algorithm follows the same elimination strategy. The Chan-Chen
algorithm eliminates the points that are not on the convex hull by traversing
the tentative hull chain in reverse order, the points with the larger 
$x$-coordinates first. This procedure is done through finding the hull
edge crossing the right wall of $\sigma$ by performing a pass over the
remaining points (those to the right of $\sigma$). Suppose a point
$P[j]$ from the remaining points is currently being inspected, by
imitating Graham's scan, the point $P[j]$ is tentatively added to the
hull if it is above the tentative hull edge found so far crossing the
right wall of $\sigma$. Also, adding $P[j]$ to the hull might require
removing some points from the chain in Graham-scan fashion. In the description
given in Figure~\ref{convex-hull-algorithm}, we do not add any new points
to the tentative hull chain. We just keep $P[j']$ as the point representing 
the right endpoint of the hull edge crossing the right wall of $\sigma$ and we 
update $j'$ accordingly. Hence, points are only removed from the hull chain 
during the elimination process. At the end of the pass, the leftover points on the chain 
are indeed on the convex hull and are accordingly reported.

Finding the next $s$ points with the smallest $x$-coordinates
requires $O(N)$ time, computing the upper hull for $s$ points requires $O(s
\lg s)$ time, and pruning the tentative hull chain from points not on the
upper hull requires $O(N)$ time. Since there are $O(N/s)$ rounds, the
algorithm runs in $O(N/s \cdot (N + s \lg s))$ worst-case time,
which is $O((N^2/S) \cdot \lg N + N \lg S)$.

\subsection{Our Algorithm}

Our convex-hull algorithm uses three navigation piles: a
(max-)augmented navigation pile that we call \MaxPile{}, and two
(min-)unaugmented navigation piles that we call \MinOne{} and
\MinTwo{}.  
Without loss of generality, we assume that the orientation (either max or min) 
of the navigation piles is with respect to the $x$-coordinates of the
points. The algorithm works as follows:

\begin{enumerate}

\item
Insert all the $N$ points in both \MinOne{} and \MinTwo{}.
\item
Let $i_0$ be the index of the point with the minimum $x$-coordinate, and $\mathbf{none}$ if the input set is empty.
\item
\textbf{while} $i_0 \neq \mathbf{none}$:

\begin{enumerate}

\item Extract the minimum $S$ points, one by one, from \MinOne{} (or
  until $|\MinOne{}| = 0$), and keep track of the first and last points. 
  These two values will be used as filters for \MaxPile{}; let us call them 
  $f_1$ and $f_2$. These $S$ points will be the candidates considered in this round, 
  determining the slab $\sigma$.

\item Reinitialize an empty \MaxPile{} using the current candidates and filters, 
  without actually inserting the points. Reinitialize the
  $\mathit{count}$ vector by scanning the $N$ points, bucket by
  bucket. Then, build the \Rank{} and \Select{} structures for the
  $\mathit{count}$ vector.

\item
\label{critical-piece}
 Construct the upper hull for the current $S$ points. This can be
  done by using a space-efficient implementation of Graham's scan. For
  example, the in-place variant described in \cite{BIKMMT04} could be
  modified to use a bit vector of alive elements, instead of swapping
  the input elements. In this computation, \MaxPile{} and \MinTwo{}
  can be deployed as follows.

\begin{enumerate}
\item Extract the two points with the minimum $x$-coordinates 
from \MinTwo{} and insert both of them into \MaxPile{}.
\item \textbf{repeat} $S-2$ times (or until $|\MinTwo{}| = 0$): 
\begin{enumerate}
\item Extract the minimum point from \MinTwo{}; let its index be $i'$.
\item \textbf{while}  $|\MaxPile{}| \geq 2$
  \textbf{and} \textbf{not} $\mathit{right}\mbox{\rm
    -}\mathit{turn}$(next to maximum of \MaxPile{},  maximum of \MaxPile{}, $P[i'])$:\newline \hspace*{1.25em}repeatedly extract the maximum point from \MaxPile{}.
\item Insert the point $P[i']$ into \MaxPile{}.
\end{enumerate}
\end{enumerate}

At this point, the alive points in \MaxPile{} form a tentative upper hull.  

\item The goal here is to eliminate the points that are not really on the hull
among the points in \MaxPile{} forming the tentative hull chain. We do so by 
computing the hull edge crossing the right wall of $\sigma$.

\begin{enumerate}
\item Set $j'$ to $\mathbf{none}$, which represents the right endpoint of the hull edge crossing the right wall of $\sigma$.
\item \textbf{for} every point with index $j$ whose $x$-coordinate is
  greater than $f_2$:

\begin{enumerate}
\item If $j' \neq \mathbf{none}$ \textbf{and} $\mathit{right}\mbox{\rm
    -}\mathit{turn}$(maximum of \MaxPile{}, $P[j']$, $P[j])$, \textbf{continue}.
\item \textbf{while} $|\MaxPile{}| \geq 2$ \textbf{and} \textbf{not} $\mathit{right}\mbox{\rm
    -}\mathit{turn}$(next to maximum of \MaxPile{},  maximum of \MaxPile{},
$P[j]$):\\ \hspace*{1.25em}
repeatedly extract the maximum point from \MaxPile{}.\item Set $j'$ to $j$.
\end{enumerate}
\end{enumerate}

\item \label{spec} Set $i_0$ to $j'$, then extract and
  neglect points from \MinOne{} and \MinTwo{} until we reach $P[i_0]$ as
  the current minimum for both.
\item  Extract all alive points from \MaxPile{} and report them as points on the convex hull, but in reverse order. This reversal can be done by inserting the points in an empty min-navigation pile, then extracting them.
\end{enumerate}
\end{enumerate}

By inspecting our algorithm and the algorithm of Chan and Chen, at
high level, the algorithms are quite similar. The main
difference is that we use adjustable navigation piles in order to
recall the points and deal with them. It should also be noted that
the full power of our techniques (augmenting the structure) is only needed 
in step~(\ref{critical-piece}). All other parts could have been handled efficiently without
augmenting the adjustable navigation pile.

\subsection{Analysis}

We need to prove that our convex-hull algorithm achieves a time
complexity of $O(N^2/S + N \lg S)$ and a space complexity of
$\Theta(S)$ bits. As the space complexity, we deploy three
navigation piles that are proved to require $\Theta(S)$ bits. As
for the time complexity, step (1) requires $O(N)$ time to build two
unaugmented navigation piles. Obviously, step (2) can
be done in $O(1)$ time. Now, we are going to analyse step (3). There
are $\lceil N/S \rceil$ rounds in total. The work done in each round
can be summed up as follows:

\begin{itemize}

\item In step (3a), extracting $S$ points requires $O(N + S \lg S)$
  time.

\item Step (3b) needs $O(N)$ time to construct the $\mathit{count}$
  vector and the data structures for answering \Rank{} and \Select{}
  queries.

\item Constructing the upper hull of $S$ points in step (3c) is done in
  $O(N + S \lg S)$ time.
Note that $S$ points are extracted from \MinTwo{}. Each of these
points will be inserted and may be extracted later from \MaxPile{}, but
a point can be inserted once and extracted once from \MaxPile{}, for a
total of at most $S$ insertions and $S$ extractions.

\item The time complexity of step (3d) is $O(N + S \lg S)$ too. We need
  $O(N)$ to loop on the whole input sequence. Also, we may be
  extracting points from \MaxPile{}. Given that the number of alive
  points in \MaxPile{} is at most $S$, then the extractions need at
  most $O(N + S \lg S)$ time.

\item Step (3e) does not affect the time complexity of the
  algorithm. Here, we extract points from \MinOne{} and
  \MinTwo{}. Since these piles initially contain the $N$ points and no
  more insertions are done into them, throughout the algorithm all
  extractions from these piles require $O(N^2/S + N \lg S)$ time.

\item Given that the number of points in \MaxPile{} is at most $S$,
  step (3f) (including the reversal of the order of the points) 
	will be done in $O(N + S \lg S)$ time.
\end{itemize}

As a conclusion, the total cost of step (\ref{spec}) is $O(N^2/S +
N\lg S)$. Except for step (\ref{spec}), the worst-case time complexity
of step (3) is $O(N + S \lg S)$ per round. Multiplying this by the
number of rounds $\lceil N/S \rceil$, the time complexity of our
convex-hull algorithm is $O(N^2/S + N \lg S)$, as claimed.

\section{Concluding Remarks}
\label{sec:remarks}

\subsection{Summary} 

When constructing adjustable navigation piles, four techniques
are important: 1)~implicit links when indexing different types of
objects, 2)~bit packing and unpacking, 3)~buffering and incremental
submersion, and 4)~quantile thinning.  In addition to their connection
to binary heaps \cite{Wil64}, we pointed out the strong connection 
between the navigation piles and tournament trees.  
Conclusively, our data structure is shown to be a useful ingredient in space-efficient algorithms 
for problems that employ incremental sorting within their engine.
In general, we illustrated some conditions for when a succinct
implementation of a priority queue that uses a workspace constituting a sublinear number of bits is possible,
so that algorithm designers would be careful when using the structure.  

On the negative side, in practice, navigation piles are slow
\cite{JK06} for two reasons: 1)~The bit-manipulation machinery is
heavy and index calculations devour clock cycles. 2)~The cache
behaviour is poor because the memory accesses lack locality.
It is expected that the situation is not better for adjustable navigation piles.

Our sorting is a heapsort algorithm \cite{Wil64} that uses
an adjustable navigation pile instead of a binary heap.  For our
convex-hull algorithm we had to augment the data structure with extra
information while still maintaining the same asymptotic memory usage.
In spite of the optimality of the asymptotic running time of our
algorithms, one could criticize the practicality of the model itself, since the
memory-access patterns may not always be friendly to contemporary
computers while it is not allowed to move the elements around.

\subsection{Related Developments}
The credit for the use of quantile thinning should go to Pagter and
Rauhe \cite{PR98}.  However, the way the technique is used in the
adjustable navigation pile leads to a simpler and more elegant data
structure.  Recently, quantile thinning has also been used in a data
structure to answer heavy-hitter queries for a set of points on a line
\cite{EHMN11}.

Buffering and incremental submersion is a general data-structural
transformation that can be used to speed up \Insert{} in priority
queues \cite{AHRT05}.  Recently, this technique has also been used in a
space-efficient manner in weak heaps \cite{EEK13} and in strengthened
lazy heaps \cite{EEK15}.

After introducing the adjustable navigation piles in the conference version of this paper, 
our data structure has also found application in space-efficient graph algorithms
\cite{EHK15} and space-efficient plane-sweep geometric algorithms \cite{EK15}.

\subsection{Other Data Structures for Read-Only Data} 
In our experience, very few data structures can be made memory
adjustable as elegantly as priority queues. A stack is a gratifying
companion \cite{BKLSS15}.  As counterweight, a dictionary must maintain a
permutation of a set of size $N$; this means that it is difficult to
manage with much less than $N \ceils{\lg{N}}$ bits. However, when the
goal is to cope with about $N$ bits, a bit vector extended with \Rank{}
and \Select{} facilities (for a survey, see~\cite{NP12}) is a
relevant data structure.  Two related constructions are the wavelet
stack used in \cite{EJKS14} and the wavelet tree introduced in
\cite{GGV03} (for a survey, see \cite{Nav12}).

\subsection{Open Problems}

The optimality of our algorithm for computing convex hulls follows
from the fact that the sorting problem reduces to the problem of
computing the convex hull of a planar point set \cite[Section
  3.2]{PS85}. However, in the restricted RAM model, for the following
variants of the problem, the exact space-time trade-offs are still
unknown:
\begin{enumerate}
\item Compute the convex hull of a planar set of points given in
  lexicographic sorted order according to their coordinates.
\item Compute the convex hull of a simple polygon. 
\item Compute the extreme points of a planar set of $N$ \emph{distinct} points,
  i.e.~the points on the convex hull in any order, and express the
  complexity as the function of $N$ and $h$, where $h$ denotes the
  size of the output.
\end{enumerate}
For these problems, the space-time lower bound obtained via the
unique-elements problem \cite[Section 10.13.7]{Sav08} does not hold any more.
For the best known results, we refer to \cite{BKLSS15,CC07} (just be
aware that in these papers the space bounds are expressed in words,
not in bits).

\subsection*{Acknowledgements}

We thank Tetsuo Asano for introducing the restricted RAM model to us
and taking part in the initial discussions on the topic that led to
the post at a conference \cite{AEK13}.

\bibliography{shortstrings,read-only-data}
\bibliographystyle{DIKU}

\end{document}